\documentclass[twocolumn,english,aps,pra,superscriptaddress,longbibliography]{revtex4-1}
\usepackage[T1]{fontenc}

\usepackage{color}
\usepackage{xcolor}
\usepackage{amsmath}
\usepackage{graphicx}
\usepackage{natbib}
\usepackage{braket}
\usepackage{amssymb}
\usepackage[bottom]{footmisc}
\usepackage[normalem]{ulem}
\usepackage[colorlinks=true,urlcolor=blue,citecolor=blue,linkcolor=blue]{hyperref}
\usepackage[ruled,vlined]{algorithm2e}
\usepackage{setspace}
\usepackage{todonotes}
\usepackage{soul}
\hypersetup{colorlinks,
	linkcolor={blue!75!black!80!yellow},
	citecolor={blue!75!black!80!yellow},
	urlcolor={blue!75!black!80!yellow}
}

\makeatletter \renewcommand\@make@capt@title[2]{%
\@ifx@empty\float@link{\@firstofone}{\expandafter\href\expandafter{\float@link}}%
\sffamily{\textbf{#1}}\@caption@fignum@sep#2 }

\usepackage[normalem]{ulem}

\makeatother

\usepackage{babel}

\begin{abstract}
    Quantum computation is conventionally performed using quantum operations acting on two-level quantum bits, or qubits. Qubits in modern quantum computers suffer from inevitable detrimental interactions with the environment that cause errors during computation, with multi-qubit operations often being a primary limitation. Most quantum devices naturally have multiple accessible energy levels beyond the lowest two traditionally used to define a qubit. Qudits offer a larger state space to store and process quantum information, reducing complexity of quantum circuits and improving efficiency of quantum algorithms. Here, we experimentally demonstrate a ternary decomposition of a multi-qubit operation on cloud-enabled fixed-frequency superconducting transmons. Specifically, we realize an order-preserving Toffoli gate consisting of four two-transmon operations, whereas the optimal order-preserving binary decomposition uses eight \texttt{CNOT}s on a linear transmon topology. Both decompositions are benchmarked via truth table fidelity where the ternary approach outperforms on most sets of transmons on \texttt{ibmq\_jakarta}, and is further benchmarked via quantum process tomography on one set of transmons to achieve an average gate fidelity of 78.00\% $\pm$ 1.93\%. 
\end{abstract}

\begin{document}

\title{Implementing a Ternary Decomposition of the Toffoli Gate on Fixed-Frequency Transmon Qutrits}
\author{Alexey Galda}
\thanks{These two authors contributed equally}
\affiliation{James Franck Institute, University of Chicago, Chicago, IL 60637, USA.}
\affiliation{Computational Science Division, Argonne National Laboratory, Lemont, IL 60439, USA}
\author{Michael Cubeddu}
\thanks{These two authors contributed equally}
\affiliation{Aliro Technologies, Inc., Brighton, MA 02135, USA}
\author{Naoki Kanazawa}
\affiliation{IBM Quantum, IBM Research Tokyo, Tokyo, 103-8510, Japan}
\author{Prineha Narang}
\email{prineha@seas.harvard.edu}
\affiliation{School of Engineering and Applied Sciences,
Harvard University, Cambridge, MA 02138, USA.}
\author{Nathan Earnest-Noble}
\email{nate@ibm.com}
\affiliation{IBM Quantum, IBM T.J. Watson Research Center, Yorktown Heights, NY 10598, USA}
\date{\today}
\date{\today}

\maketitle

\section{Introduction}
With the accelerating interest in near-term practical quantum advantage, there is a push to improve the performance of short-depth quantum circuits. Conventionally, these circuits are decomposed into a set of native instructions, or basis gates, limited to operations on two-level quantum systems. However, most quantum devices naturally have multiple accessible energy levels beyond the lowest two traditionally used to define a qubit. By extending the available set of basis gates to include qudit-based operations, one can achieve several advantages, such as alternate qubit gates~\cite{baksic2016speeding,zhou2017accelerated,earnest2018realization,abdumalikov496experimental, egger2019entanglement,yurtalan2020implementation,hazard2019nanowire,gyenis2019experimental}, improved qubit readout~\cite{jurcevic2021demonstration,mallet2009single}, higher encoding capabilities~\cite{morvan2020qutrit,blok2020quantum,cervera2021experimental,wang2020qudits,gokhale2020extending,kues2017chip,neeley2009emulation}, and more efficient qubit circuit decompositions~\cite{ralph2007efficient}, such as with qutrits and the Toffoli gate~\cite{ralph2007efficient, gokhale2019asymptotic,fedorov2012implementation,hill2021realization,baekkegaard2019realization,baker2020improved}. 

The Toffoli gate, a controlled-controlled-\texttt{NOT} (\texttt{CCNOT}), and its $n$-qubit generalization~\cite{ralph2007efficient} is of great interest in quantum computing as reversible multi-qubit gates play a crucial role in a number of quantum computing applications such as quantum error correction schemes~\cite{cory1998experimental, schindler2011experimental}, fault-tolerant quantum computing~\cite{dennis2001toward, paetznick2013universal}, and quantum search algorithms~\cite{shor1995scheme, hu2019efficient}. While it has been shown that the optimal binary decomposition of the quantum \texttt{CCNOT} gate requires five general two-qubit gates~\cite{divincenzo1998quantum}, when we constrain the native gate set to include only \texttt{CNOT}s and arbitrary single-qubit gates, the optimal and standard decomposition of \texttt{CCNOT} requires six \texttt{CNOT} gates~\cite{margolus1994simple}. In the case of the heavy-hex lattice configuration~\cite{chamberland2020topological} architecture employed by IBM Quantum - the most efficient decomposition can be achieved with seven \texttt{CNOT}s, or eight \texttt{CNOT}s if we additionally constrain the initial logical qubit order to be preserved~\cite{toffoli_lnn}. By leveraging intermediate ternary quantum logic with qutrit gates, we achieve asymptotic improvements to entangling gate counts relative to qubit-based decompositions of \texttt{CCNOT} and its multi-controlled generalizations~\cite{gokhale2019asymptotic,ralph2007efficient}. Using Qiskit Pulse~\cite{alexander2020qiskit}, we can efficiently expand the existing native gate set of IBM Quantum's~\cite{IBMQuantum} superconducting transmon devices to include single-qutrit and two-qutrit gates, allowing users to experimentally implement shorter-depth decompositions of the \texttt{CCNOT} gate using fewer entangling operations, with minimal calibration overhead.

In this work, we demonstrate that the extension to ternary quantum logic provides a path for reducing circuit depth of quantum circuits that rely on \texttt{CCNOT} gates, with an emphasis on experimentally implementing two-qutrit gates with the cross-resonance interaction~\cite{sheldon2016procedure} on fixed-frequency transmons~\cite{koch2007charge}. In doing so, we reveal some of the functional behavior arising from the cross-resonance interaction between qutrits, and demonstrate an optimization on cloud-enabled near-term quantum hardware by extending the framework of quantum computation to encompass ternary quantum logic. 
Our work is distinct from previous ternary decompositions of \texttt{CCNOT} gates~\cite{fedorov2012implementation,hill2021realization} due to its experimental implementation on a fixed-frequency transmon architecture, the use of newly-developed techniques~\cite{stenger2021simulating} to expand the basis gate set with no additional calibration experiments, as well as the development of a just-in-time error mitigation technique~\cite{wilson2020just} which is a software-compatible approach to reduce charge noise induced errors.

The remainder of this paper is structured as follows: in Section II, we introduce the \texttt{CCNOT} gate and its decomposition on devices with the heavy-hex topology, highlighting the necessary gates for a qutrit-based implementation. Section III discusses the experimental results, demonstrating limitations arising from charge noise and our methods to mitigate its impact. We then end with conclusions in Section IV, and a discussion of the impact of this work. 
\begin{figure}[t!]
\centering
\includegraphics[width=\columnwidth]{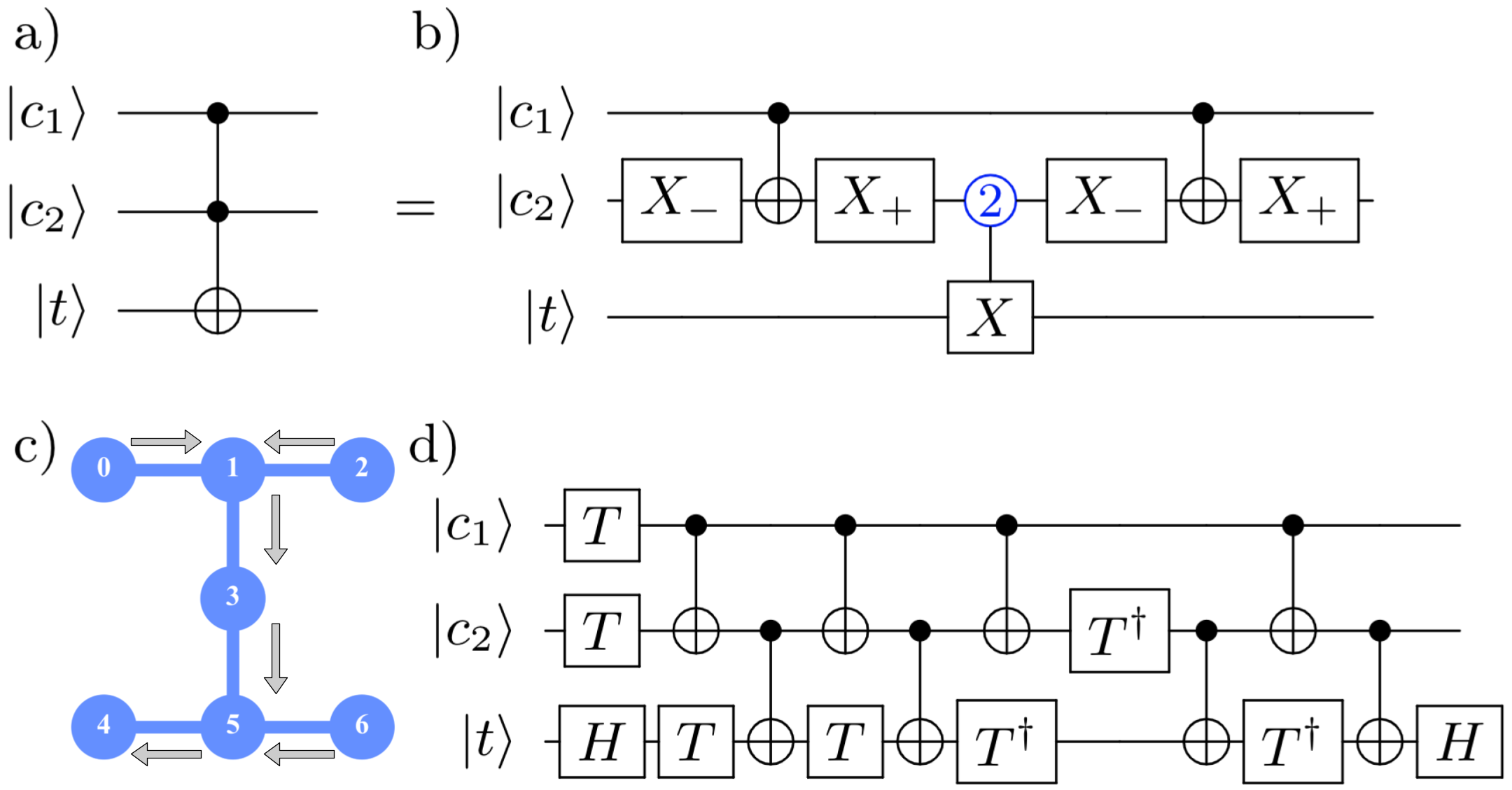}
\caption{Circuit diagram representation of a) \texttt{CCNOT} gate and b) its decomposition using single- and two-qutrit gates. Only one transmon (control $c_2$) is a qutrit, with the other control and target transmons remaining in the qubit subspace. c) Physical topology of \texttt{ibmq\_jakarta}, with the arrows indicating the native CNOT direction that is calibrated. d) Order-preserving qubit-based \texttt{CCNOT} gate decomposition using single-qubit gates and eight \texttt{CNOT}s on the heavy-hex topology~\cite{toffoli_lnn}.}
\label{fig:decomps}
\end{figure}
\section{Toffoli Decomposition and Experimental Gates}

\begin{figure*}[t!]
\centering
\includegraphics[width=\textwidth]{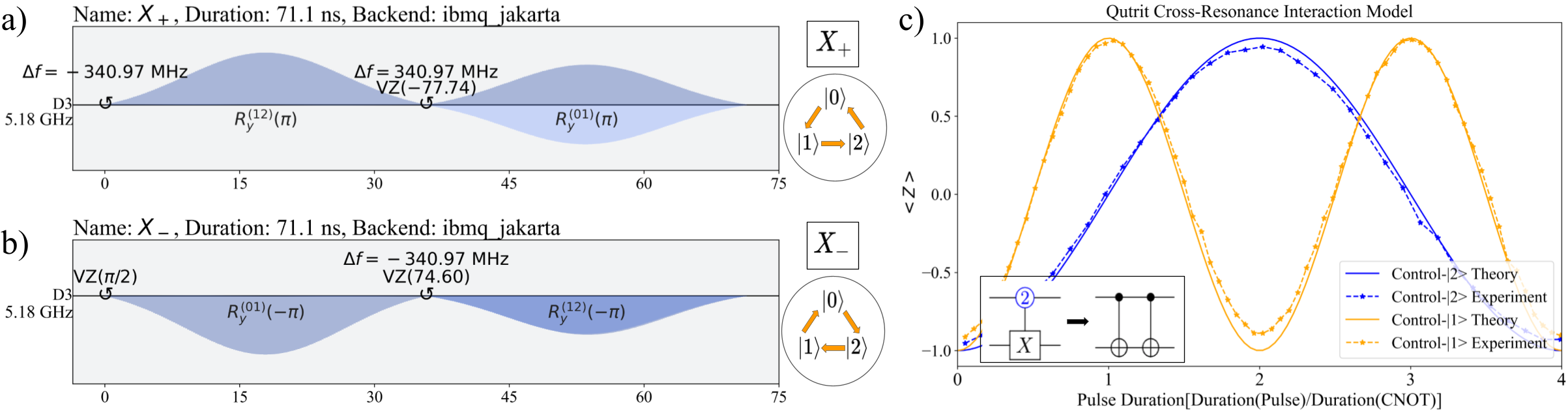}
\caption{a) and b) The $X_+$ and $X_-$ pulse schedules and the corresponding gate on the right describing its state transitions. Pulses are played on the drive channel of transmon 3 of \texttt{ibmq\_jakarta} (\texttt{D3}) with the phase shift implemented by the virtual-Z gate and frequency shift represented by \texttt{VZ} and $\Delta f$ with its amount, respectively.
c) Dynamics of the standard echoed cross resonance pulse under different control transmon states. 
Dots are the experimental results which use the scaling technique presented in ~\cite{stenger2021simulating}, and bold lines are the model calculation described in the main text.
The horizontal axis shows the pulse duration normalized by the cross resonance duration of the device calibrated \texttt{CNOT} gate.
The inset shows the decomposition of the $\ket{2}$-controlled-\texttt{NOT} gate.}
\label{fig:qutritgates}
\end{figure*}

To experimentally implement our ternary \texttt{CCNOT} decomposition, we explore modifications to previous theoretical proposals~\cite{ralph2007efficient, lanyon2009simplifying, gokhale2019asymptotic}, taking into account the experimental interaction of the cross-resonance gates available on the IBM Quantum superconducting quantum processor \texttt{ibmq\_jakarta}, shown in Fig.~\ref{fig:decomps}c. The corresponding quantum circuit consists of single-qutrit gates $X_{\pm}$, defined as $X_{\pm}\ket{i} = \ket{(i\pm1)\textnormal{ mod }3}$. 
This quantum operation requires different transition energies depending on $\ket{i} \in \{\ket{0}, \ket{1}, \ket{2}\}$ in the anharmonic oscillator system, and one cannot coherently drive all of these transitions with a single excitation pulse.
Thus, $X_{\pm}$ gates employ the following experimental realization using a combination of coherent excitations on different energy subspaces as shown in Fig.~\ref{fig:qutritgates}a and b, namely the transitions between $\ket{0}$ and $\ket{1}$ at frequency $\omega_{01}$, and $\ket{1}$ and $\ket{2}$ at frequency $\omega_{12}$:
\begin{align}
    X_{+} &= R_y^{(01)}(\pi)R_y^{(12)}(\pi)\,,\\
    X_{-} &= R_y^{(12)}(-\pi)R_y^{(01)}(-\pi)\,,
\end{align}
where $R_\beta^{(ij)}(\theta)$ is a unitary operation represented by $\exp(-i\frac{\theta}{2} \sigma_\beta^{(ij)})$, the $\sigma_\beta^{(ij)}$ is a generalized Pauli operator, $\beta \in \{ x, y, z \}$ is the operational Pauli basis of the truncated three-dimensional Hilbert space and $ij \in \{0, 1, 2\}^2$ represents a pair of excitation levels to operate.

Furthermore, standard \texttt{CNOT} gates between control qubits $c_1$ and $c_2$, and a $\ket{2}$-controlled \texttt{NOT} gate between control qutrit $c_2$ and target qubit $t$, allow us to achieve the ternary decomposition for a \texttt{CCNOT} gate shown in Fig.\ref{fig:decomps}b. The $\ket{2}$-controlled \texttt{NOT} gate is defined as inverting the state of the target qubit if and only if the control is in the $\ket{2}$ state, which we achieve by constructing a composite two-qutrit gate as shown in Fig.~\ref{fig:qutritgates}c (inset), bringing the total two-transmon gate count to four \footnote{The device implementation of the $\ket{2}$-controlled \texttt{NOT} gate required cross-resonance pulses of double the duration of those for a \texttt{CNOT} gate, when using default gates. It may be possible to implement an alternative tuning method to achieve a $\ket{2}$-controlled \texttt{NOT} which is more efficient}.

One key advantage of this ternary decomposition of the \texttt{CCNOT} gate is that it can be realized on the heavy-hex architecture and does not require all three transmons to be physically connected. More specifically, the ternary decomposition is inherently ``order-preserving'', as it does not require the insertion of any additional \texttt{SWAP} gates, provided the target qubit is allocated to one end, and control qutrit $c_2$ is allocated to the middle transmon. The analogous depth-optimal, order-preserving, qubit-based decomposition of the \texttt{CCNOT} gate on the heavy-hex architecture requires eight \texttt{CNOT} gates~\cite{toffoli_lnn}, as shown in Fig.~\ref{fig:decomps}d. 

An important consideration when carrying out qutrit gates is the relative phase precession between different qutrit subspaces originating from the transmon anharmonicity. When a pulse is applied in the $(01)$ subspace, the phase of quantum states in the $(12)$ qutrit subspace experiences a precession with anharmonic frequency $\alpha = \omega_{12} - \omega_{01}$. As a result, tracking of the relative phase between qutrit subspaces has to be actively implemented in order to account for this precession and to retain the desired phase in all subspaces. The details of accounting for this phase precession can vary depending on the underlying qubit architecture and the available control and drive channels. At the time of writing, qutrit phase tracking for IBM quantum processors is not currently implemented in Qiskit Pulse and was thus implemented by the authors.

For the required two-qutrit gate, $\ket{2}$-controlled \texttt{NOT}, we leverage the existing all-microwave-echoed cross-resonance gate available on IBM transmon devices. While this gate has been studied extensively in the context of qubits~\cite{sheldon2016procedure,malekakhlagh2020first,sundaresan2020reducing}, generally these studies restrict the resulting interaction analysis to the qubit subspace. 
In this work, we focus on the dynamics of the target transmon when the control transmon stays in the $\ket{2}$ state, and analytically and experimentally investigate such dynamics. Firstly, we prepare the control transmon $c_2$ in either the $\ket{1}$ or $\ket{2}$ state, then apply the echoed cross-resonance (ECR) tone that is resonant to $\omega_{01}$ of the target transmon. Each cross-resonance tone forms a flat-topped pulse with Gaussian rising and falling edges with $\sigma$ fixed at 14.08 ns. We vary the total gate duration of the ECR tone using a calibration-free scaling technique~\cite{stenger2021simulating}  and measure the target qubit in the $Z$ basis to extract its expectation value. 
As shown in Fig.~\ref{fig:qutritgates}c, the controlled rotation rate is halved when the control qubit stays in $\ket{2}$ state.
These dynamics may be interpreted by the naive extension of the cross-resonance Hamiltonian \cite{Magesan_2020} in which unwanted interactions from higher levels are ignored for the sake of simplicity, $ZI$ and $IX$ terms are eliminated by the echo, and under the assumption of the time-invariant microwave drive
\begin{align}
    U_{ZX}(\tau) = \exp\left(-i\Omega_{ZX}^{(01)}\frac{\sigma_Z^{(01)} \otimes \sigma_X^{(01)}}{2}\tau\right),
\end{align}
where the interaction term of interest is redefined in the two-qutrit subspace. We calculate the same expectation value at each gate time $\tau$ with different initial states. This calculation agrees well with the experimental result, indicating that the state-dependent interaction rate arises from the $ZX$ term of the cross-resonance Hamiltonian.
Thus, we find the $\ket{2}$-controlled \texttt{NOT} gate can be implemented with two consecutive \texttt{CNOT} gates, without any modification of the pulse schedule.
In principle, one can implement the same gate dynamics with a single ECR with the high-power drive, however, this is not practical due to saturation of the interaction rate as well as leakage to higher energy levels.
Note that the result of Fig.~\ref{fig:qutritgates}c is insensitive to the phase accumulation in each energy level, which should be corrected for in order to implement the decomposition of Fig.~\ref{fig:decomps}b.

\section{Experimental Implementation of Toffoli Gate}
\begin{figure}[tb]
\centering
\includegraphics[width=0.77\columnwidth]{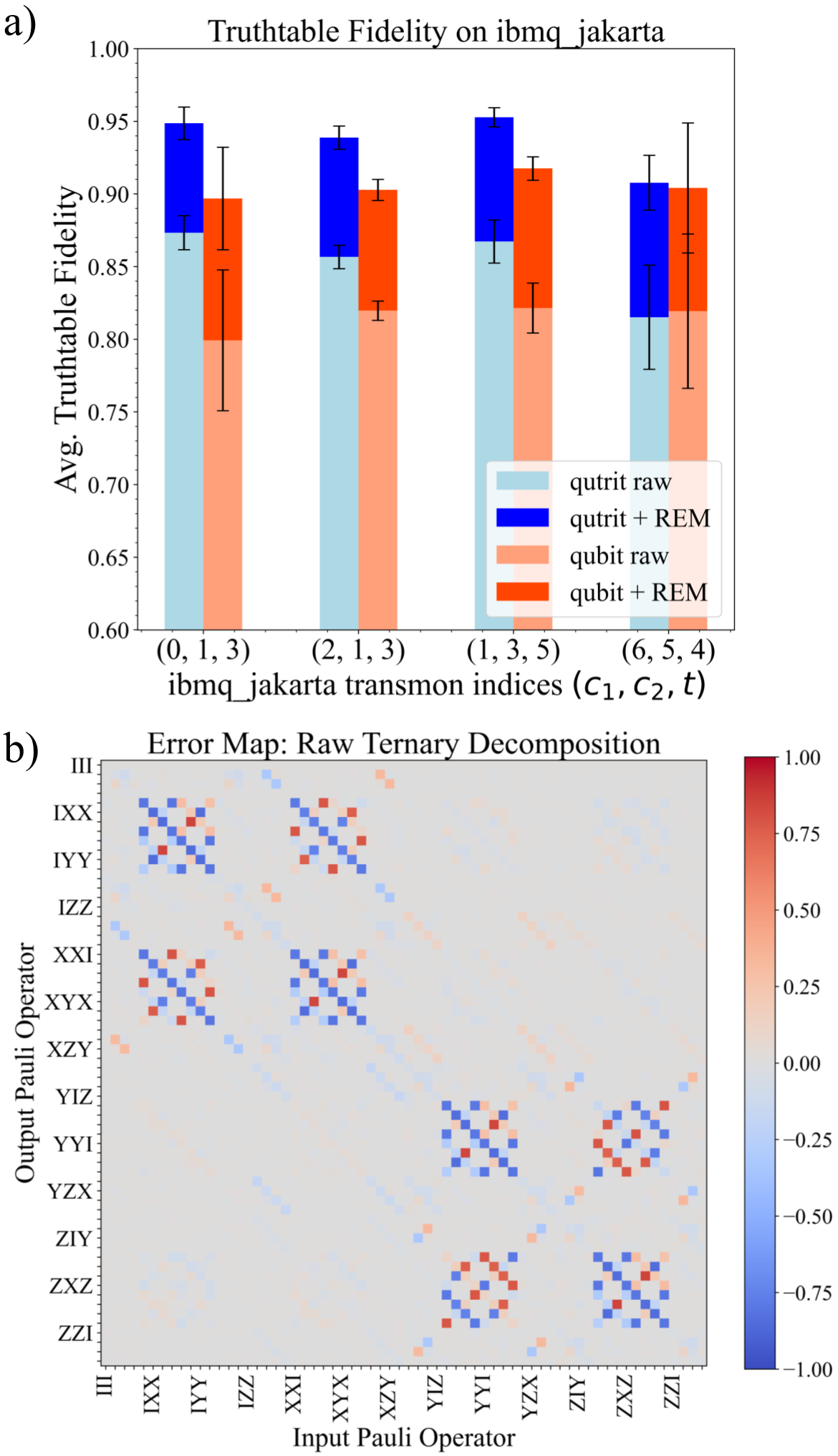}
\caption{a) Truth table fidelities $\mathcal{F}_{TT}$ for the ternary \texttt{CCNOT} decomposition benchmarked against the analogous qubit-based decomposition on \texttt{ibmq\_jakarta} with and without REM. b) Detailed gate error of a specific experiment with transmons (0, 1, 3) of \texttt{ibmq\_jakarta}. The gate error is visualized in the form of difference of the Pauli transfer matrix (PTM) from the ideal PTM of the \texttt{CCNOT} gate.}
\label{fig:results}
\end{figure}

Using the developments discussed in the previous section, we then implemented the full \texttt{CCNOT} gate using our proposed ternary decomposition. We began benchmarking our ternary gates on all eight classical computational state inputs $\ket{ijk}\forall i,j,k\in\{0, 1\}^3$ to get an overall truth table fidelity as defined by  $\mathcal{F}_{TT} = (1/8)\mathrm{Tr} [U^\dag_{\mathrm{exp}} U_{\mathrm{ideal}}]$ for classical input states, with the corresponding results shown in Fig.~\ref{fig:results}a. This simple fidelity benchmark is a good first step in gate characterization and verification, because it just consists of $2^3 = 8$ different circuit inputs. We test this across four three-transmon subsystems of the backend \texttt{ibmq\_jakarta}, with and without usage of the tensor product readout error mitigation (REM) \cite{Bravyi_2021}, and find that the ternary decomposition outperforms the best binary decomposition, giving a mean improvement of $\mathcal{F}_{TT}$ by 3.82\% without REM and 3.16\% with REM, when averaged over all benchmarked three-transmon subsystems of \texttt{ibmq\_jakarta}. Furthermore, the total gate time of the ternary decomposition used in these experiments was 1.593 $\mu$s, compared to the gate time of the binary decomposition of 2.510 $\mu$s.
This result indicates two-qubit gate number reduction in the \texttt{CCNOT} decomposition can improve the gate performance.

For a more holistic characterization of our gate implementation, one should use quantum process tomography (QPT) and verify the average gate fidelity $F_{\rm avg}$.
The process is measured with the input basis $\{\ket{0}, \ket{1}, \ket{+}, \ket{+i}\}$ and measurement basis $\{X, Y, Z\}$ for each qubit, namely, $4^3 \times 3^3 = 1728$ circuits are used to reconstruct the quantum process of the \texttt{CCNOT} gate.
We perform 1024 repetitions for each QPT basis configuration and 2048 repetitions for the REM calibration routine. The quantum channel is tomographically reconstructed using the convex maximum likelihood estimation method with the completely positive and trace-preserving constraints.
After conducting several QPT experiments on the same qutrit decomposition that gave improved average $\mathcal{F}_{TT}$, we observed the more holistic average gate fidelity $\mathcal{F}_{\rm avg}$ (calculated from QPT in conjunction with REM over several three-transmon subsystems and days) can range from as low as 18.62\% to as high as 78.75\%. This instability is likely caused by the unexpected phase accumulation by random telegraph noise (RTN) \cite{Faoro_2004} especially in the qutrit frame during the $\ket{2}$-controlled \texttt{NOT} operation.
Note that the truth table benchmarks are insensitive to such phase noise because these experiments are always performed with the control transmon $c_2$ in the eigenstate of $\sigma_Z^{(12)}$ or $\sigma_Z^{(01)}$.
The reconstructed process matrix shown in Fig.~\ref{fig:results}b also indicates a significant non-local error, which cannot be easily corrected by any post single-qubit rotations.

\begin{figure*}[t!]
\centering
\includegraphics[width=\textwidth]{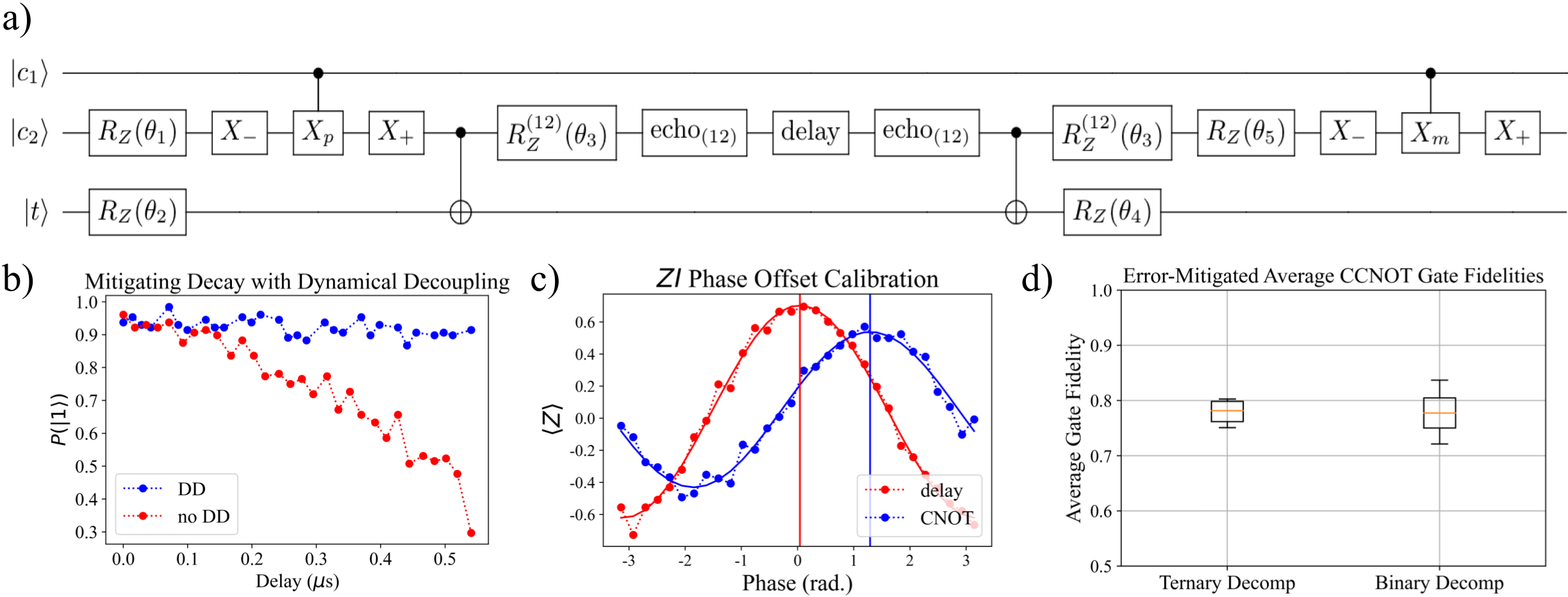}
\caption{a) The \texttt{CCNOT} gate decomposition with correction for unwanted phase accumulation. $R_Z^{(12)}(\theta)$ corrects for the phase imparted by the cross resonance drive, whereas $R_Z(\theta)$ is a local phase correction to maximize the measured process fidelity as shown in the Ref~\cite{alexander2020qiskit}. These phase corrections are implemented by the virtual-$Z$ gate with zero duration, and values are determined by the just-in-time calibration procedure prior to the execution of a QPT experiment. The delay in the middle has the twice duration of the device calibrated \texttt{CNOT} gate. b) The impact of dynamical decoupling in the qutrit subspace confirmed by the conventional Ramsey experiment. Red (blue) curve is measured without (with) the dynamical decoupling. No drive is applied during the delay. c) The calibration of the AC-Stark shift -induced phase accumulation in the $\ket{2}$ state. Red (blue) curve indicates the Ramsey experiment with various phase rotations without (with) the ECR drive. The maximum point indicates the phase that cancels out the accumulated phase. d) Box plots of the ternary decomposition average gate fidelity as determined by QPT, as well as the most efficient qubit-based implementation.}
\label{fig:errormitresults}
\end{figure*}

In order to implement a stable, high-fidelity decomposition, we introduce a just-in-time software-based method to mitigate the observed phase uncertainty by incorporating the dynamical decoupling (DD) technique~\cite{pokharel2018demonstration,viola1998dynamical,Faoro_2004,duan1999suppressing,zanardi1999symmetrizing,viola1999dynamical} in the (12) subspace, and appending phase corrections to different parts of the circuit as shown in Fig. \ref{fig:errormitresults}a and further described in subsequent text.
This technique is known to be effective to improve conventional circuit decompositions in which an even number of $X$ gates are inserted onto idle qubits, instead of delays, to protect them from the RTN \cite{jurcevic2021demonstration}.
In the frame of the qubit, this sequence can be seen as $\tau/4 \cdot X^{(12)} \cdot \tau/2 \cdot X^{(12)} \cdot \tau/4$, where $\tau$ indicates the delay in between instructions. In this sequence, an arbitrary unknown phase offset is accumulated for $\tau/2$ with the opposite sign and is thus cancelled out.
The efficacy of this method is confirmed with the Ramsey experiment shown in Fig.~\ref{fig:errormitresults}b.
In this experiment, we prepare the transmon $c_2$ in $(\ket{0}+\ket{2})/2$ and measure in the same basis after the total delay of $\tau$.
Without using DD, the input quantum state rapidly decays due to the phase noise, whereas with DD the state is well-protected for sufficient time to play two device-calibrated \texttt{CNOT} gates with gate times of 341 ns each for \texttt{ibmq\_jakarta} transmons 3 and 5.

However, the implementation of DD simultaneously with the $\ket{2}$-controlled \texttt{NOT} is not as straightforward as the above scenario due to the non-negligible AC-Stark shift arising from the off-resonant driving of the control transmon $c_2$ \cite{Magesan_2020}.
Here, we replace the first and last delay with the device calibrated \texttt{CNOT} gates, but we must aim to keep the appropriate amount of phase accumulation to implement DD.
We first calibrate the amount of extra phase accumulation in the $\ket{2}$ state imparted by the cross resonance drive.
This is experimentally obtained by a different Ramsey experiment, in which the transmon $c_2$ is prepared in the same state but rotated by $R^{(12)}_Z(\theta_{ZI})$ for $\theta_{ZI} \in [-\pi, \pi]$ before the measurement.
The delay time is set to the gate time of the default \texttt{CNOT} gate and the measured population peaks at the $\theta_{ZI}$ that cancels the extra phase accumulation.
As shown in Fig.~\ref{fig:errormitresults}c, the population measured after the delay peaks at $\theta_{ZI, 0} = 0.1083$ rad., whereas one after the \texttt{CNOT} gate shows $\theta_{ZI, 1} = 1.1916$ rad.
The difference $\Delta \theta = \theta_{ZI, 1} - \theta_{ZI, 0}$ indicates the phase imparted by the Stark shift, which is further corrected by the virtual $R^{(12)}_Z$ gates~\cite{mckay2017efficient} in Fig.~\ref{fig:errormitresults}a.

While the Toffoli gate decomposition shown in Fig.~\ref{fig:decomps}b calls for two \texttt{CNOT} gates between control qubits $c_1$ and $c_2$, they can be replaced by a pair of controlled-$X_p$ and controlled-$X_m$ gates, where $X_{p,m} \equiv R_x(\pm \pi)$, as shown in Fig.~\ref{fig:errormitresults}a, without affecting the result. It is important to note that while in $SU(2)$ the $X$ and $X_{p,m}$ gates are equivalent because the $\pm i$ prefactor can be absorbed into an immeasurable global phase. For qutrit gates in $SU(3)$, this is not the case and extra care needs to be exercised when generalizing qubit gates, such as $X$ or $\texttt{CNOT}$, to qutrit systems. For the purpose of implementing the high-fidelity decomposition, we used the unitarily-equivalent variant with controlled- $X_p$ and $X_m$ gates, rather than \texttt{CNOT}s between the transmons $c_1$ and $c_2$.

By combining these techniques, we arrive at our final experimental ternary decomposition shown in Fig.~\ref{fig:errormitresults}a, from which we can achieve a relatively good average gate fidelity as shown in Fig.~\ref{fig:errormitresults}d where the fidelities of the ternary decomposition end up being comparable to the qubit-based implementation, albeit with less variance.
We also confirmed the effectiveness of this technique with other transmon subsystems as well as with another device available in the IBM Quantum service.
In the current implementation, achievable $\mathcal{F}_{\rm avg}$ seems to be limited by the total gate time, notably due to the fact that the employed DD technique doubles the gate time of the the $\ket{2}$-controlled \texttt{NOT} in exchange for phase stability. As a consequence, the total gate time of the phase-stabilized ternary decomposition becomes 2.432 $\mu$s whereas the eight-\texttt{CNOT} decomposition is 2.510 $\mu$s, despite the halving of the \texttt{CNOT} gate count. This indicates there is ample room for improving the phase error correction techniques either via improved software methods or hardware adjustments. For example, increasing the $E_j/E_c$ ratio of the Josephson junction ($\sim 35$ for devices used in this work) can reduce sensitivity to charge noise \cite{Blais_2021}, with recent qutrit-based work demonstrating higher fidelity qutrit-based operations for $E_j/E_C \sim 70$~\cite{blok2020quantum,morvan2020qutrit}.

The correction technique we introduce in this work is readily available in many qutrit realizations, not limited to superconducting devices, as long as the noise characteristic time is sufficiently longer than the gate time of a \texttt{CNOT}.
Recently, an alternate way of implementing \texttt{CCNOT}-like decompositions for the fixed-frequency transmon system was proposed \cite{kim2021highfidelity}, in which all three qubits were simultaneously driven to implement the Hamiltonian in the form of $ZIX + IZX$, which effectively implements a three-transmon operation.
It is demonstrated that this Hamiltonian realizes the fast and high-fidelity \texttt{CCNOT} operation at the cost of extra calibration overhead.
On the other hand, we demonstrate that the $\ket{2}$-controlled \texttt{NOT} gate can be implemented with two consecutive standard \texttt{CNOT} gates, which are contained in the native instruction sets of many superconducting devices. The fundamental qutrit pulses were also introduced in some recent quantum computing systems for the excited state promotion readout technique \cite{jurcevic2021demonstration}.
That is to say, our decomposition is readily available by combining existing system resources.
The gate duration of the phase-corrected ternary decomposition can be further reduced by adopting the direct \texttt{CNOT} gate instead of the ECR realization \cite{wei2021quantum}.
These advantages may drastically benefit the scalability of quantum processors.

\section{Conclusion}
In conclusion, we experimentally implemented and benchmarked the \texttt{CCNOT} gate decomposition by leveraging qutrit gates on fixed-frequency transmons and modifications to the existing ECR gate, neither requiring further calibration experiments to achieve truth table fidelity that outperforms the default approach by considerable margins.
Moreover, by introducing the dynamical decoupling technique to overcome the phase instability of higher-energy states and our own just-in-time error mitigation technique to correct for local phase offsets, we achieved an stable QPT gate fidelity witgh an average of 78.00\% $\pm$ 1.93\% on IBM Quantum backend \texttt{ibmq\_jakarta} (transmons 6,5,4) - which is comparable with the most efficient qubit decomposition, albeit with considerably less variance. Since most of the key ingredients to realize our decomposition are readily available in the today's standard quantum computing systems, this decomposition offers the benefit of efficient execution of the \texttt{CCNOT} gate with minimum calibration overhead. Further fidelity improvements can be achieved by careful engineering of the hardware itself to reduce the charge noise sensitivity of the higher-energy levels comprising the qutrit computational subspace. Given that these systems are not optimized for qutrit operation yet still result in comparable fidelities, this work demonstrates the promise of qutrit-based computation. Future work may include reduced overhead for error mitigation experiments, implementations on qutrit-optimized hardware, and experimental realization of more general qudit-based gates for system-efficient multi-controlled gate decompositions.

\section{Acknowledgments}
The authors acknowledge use of IBM Quantum services for this work, and would like to thank D.J. Egger, D.C. McKay, J.M. Chow, and A. Javadi-Abhari for useful conversations, and N.T. Bronn and T.L. Scholten for a careful read of the manuscript. Work by M.C. and P.N. is primarily supported by Air Force STTR grant numbers FA8750-20-P-1721 and FA8750-20-P-1704. P.N. further acknowledges support from the Quantum Science Center (QSC), a National Quantum Information Science Research Center of the U.S. Department of Energy (DOE).

\bibliographystyle{unsrt}
\bibliography{references}

\end{document}